# *APOLLO: A GPT-based tool to detect phishing emails and generate explanations that warn users*

Giuseppe Desolda

University of Bari "A. Moro", Italy, giuseppe.desolda@uniba.it

Francesco Greco

University of Bari "A. Moro", Italy, francesco.greco@uniba.it

Luca Viganò

King's College London, UK, luca.vigano@kcl.ac.uk

Phishing is one of the most prolific cybercriminal activities, with attacks becoming increasingly sophisticated. It is, therefore, imperative to explore novel technologies to improve user protection across both technical and human dimensions. Large Language Models (LLMs) offer significant promise for text processing in various domains, but their use for defense against phishing attacks still remains scarcely explored. In this paper, we present *APOLLO*, a tool based on OpenAI's GPT-4o to detect phishing emails and generate explanation messages to users about why a specific email is dangerous, thus improving their decision-making capabilities. We have evaluated the performance of APOLLO in classifying phishing emails; the results show that the LLM models have exemplary capabilities in classifying phishing emails (97% accuracy in the case of GPT-4o) and that this performance can be further improved by integrating data from third-party services, resulting in a near-perfect classification rate (99% accuracy). To assess the perception of the explanations generated by this tool, we also conducted a study with 20 participants, comparing four different explanations presented as phishing warnings. We compared the LLM-generated explanations to four baselines: a manually crafted warning, and warnings from Chrome, Firefox, and Edge browsers. The results show that not only the LLM-generated explanations were perceived as high quality, but also that they can be more understandable, interesting, and trustworthy than the baselines. These findings suggest that using LLMs as a defense against phishing is a very promising approach, with APOLLO representing a proof of concept in this research direction.

## 1 INTRODUCTION

In a world that is becoming more and more dependent on its digital counterpart, phishing attacks pose a substantial risk to users, organizations, and IT systems. Phishing is indeed one of the most used attack vectors for a variety of purposes, like stealing credentials and spreading malware [37]. These attacks are mainly effective because they leverage human vulnerabilities that can be found potentially in every user, such as their lack of knowledge, stress, and lack of time [23]. Therefore, as phishing attacks continue to get more complex and sophisticated, there is an urgent need to enhance the effectiveness of phishing protections, addressing both technological cybersecurity solutions [2, 39] and human aspects of security [23, 46].

One of the main defense mechanisms is the use of warning dialogs (which we will refer to as simply "warnings"), which alert users about potential threats when their systems (e.g., browsers or email clients) detect suspicious content. Warnings leave the final decision to the user about whether to access or not an email or website that is flagged as suspicious. Nonetheless, phishing attacks are still very effective [36], which demonstrates that, despite the precise technical countermeasures for detecting suspicious content [2, 39], the existing warnings still fail to protect users adequately. Recent works attempted to improve conventional warnings by adding explanations that can help users in the decision-making process, gain trust in the system and be motivated to heed the warning [10, 13, 64]. The benefits of warnings that include explanations have been recently demonstrated in studies such as [13, 21]; however, producing explanations is currently a manual process that takes time and human effort, being also error-prone and not easily scalable over time to adapt to new phishing attacks.

An intriguing avenue in cybersecurity is leveraging Large Language Models (LLMs) as a proactive defense tool against phishing attacks. LLMs such as OpenAI's GPT have showcased remarkable abilities in detecting patterns and anomalies in emails or websites indicative of potential phishing attempts with surprising accuracy [33, 41]. Moreover, the prospect of using LLMs to generate warning messages and explain the model's suspicions could significantly enhance protection against phishing threats. This approach, being faster than the manual design of explanations [21], would dynamically produce personalized explanations that vary based on the phishing content.

To investigate the potential of LLMs in detecting phishing emails and address the limitations of existing warnings, we created APOLLO (*Advanced Phishing preventiOn with Large Language model-based Oracle*), a tool based on GPT-4o for the automatic classification of phishing emails and for the generation of explanation messages that describe the reasons why an email is suspicious and should not be trusted. The designation "APOLLO" was derived from the Greek deity Apollo, the god of prophecy. The analogy is based on the capability of APOLLO to make predictions about emails alerting users in case of phishing attempts. The tool, which is shown in Figure 1, was developed in Python and interacts with OpenAI's APIs to use their latest model, GPT-4o [52] (at the time of writing this article). To assess the tool's capabilities in detecting phishing emails, we performed a thorough evaluation process that also considered challenging scenarios in order to measure the performance of GPT-4o as is, augmented with external information [43], and primed toward erroneous classifications. The results of this evaluation showed that GPT-4o performs very well in classifying phishing emails and that its performance can be improved by integrating information from external services, specifically URL information collected by VirusTotal API, which allowed us to achieve 99% accuracy. However, incorrect information from these services could lead to misclassification of emails (accuracy less than 0.47%), especially in legitimate emails. Fortunately, the classification outcome of phishing emails is not strongly affected, as GPT-4o adopts a safe behavior by default and ensures a high recall in detecting threats.

We also conducted an exploratory user study with 20 participants to investigate *how users perceive explanations in warnings for phishing attacks generated by LLMs*. To also have a preliminary evaluation against human-generated warnings, we included a comparison between LLM-generated warnings and state-of-the-art solutions, which comprise a warning with manually created explanations and the warning provided by the browsers Google Chrome, Mozilla Firefox, and Microsoft Edge. The study results highlighted the potential benefits of creating explanations that are more understandable, interesting, and credible than those at the state-of-the-art.

This paper continues as follows: Section 2 discusses related work on warning dialogs and the use of LLMs as a defensive tool for phishing. In Section 3, we present APOLLO, the GPT-based tool for classifying phishing emails and generating explanations automatically. Section 4 reports the evaluation process of APOLLO using GPT-4o to assess the tool's performance in classifying phishing emails. Section 5 details the user study conducted to investigate how users perceive explanations generated by APOLLO. Section 6 presents the results of the user study and compares them with four baselines [21]. In Section 7, we discuss the lessons learned from the results of the user study. Section 8 concludes the paper and presents some possible future work for this research.

## 2 RELATED WORK

### 2.1 Warning dialogs to protect users from phishing content

Eliminating the threat of phishing is currently not feasible with automated methods, as there are no existing tools that detect phishing content, e.g., websites or email, with 100% accuracy [26, 29]. Moreover, aiming for a higher recall has the side effect of lowering the precision of the content classification task [57]. Trying to minimize the number of false negatives (i.e., undetected phishing emails) would increase the number of false positives (i.e., genuine emails classified as phishing ones); this would risk jeopardizing users' productivity, as they would eventually lose genuine emails due to misclassification. To limit the number of false negatives, automated solutions usually apply a first filter based on systems like blocklists [31]. Then, contents in a "grey zone" (e.g., classified with a 60% to 95% probability of being phishing) are instead shown together with a warning dialog that alerts users of possible dangers, leaving the final decision up to them [42]. However, most users do not understand the importance and the meaning of these warnings, so this type of attack remains one of the most effective and widespread [36].

Different problems limit the effectiveness of warnings in protecting users from phishing attacks. The first one is the *passive behavior*. Early warnings consisted of indicators on the screen and suspicious content, such as security toolbars. These are called *passive warnings,* and the results are highly ineffective since users often ignore them as they fail to be even noticed [25, 72].[25, 72]. Despite this being a consolidated result from more than 10 years, passive warnings are today the most adopted solution for email clients. On the other hand, warnings that interrupt the interaction flow and force the user's attention are called *active warnings* and are more effective than passive warnings [13, 25, 54, 69].[13, 25, 54, 69]. Today, active warnings are implemented only in web browsers [22].

Another problem that limits warnings for phishing attacks is the so-called habituation effect, which happens when users are exposed to the same warning (the stimulus), even under different risk circumstances [1]. This effect causes the warning effectiveness to greatly decrease, as it leads to a reduced attentional response in users, who are more likely to ignore it [40]. To solve this problem, *polymorphic* warnings can be adopted, i.e., they change their aspect and/or content to provide different visual stimuli for the user. A study by Anderson et al. [4] demonstrated that polymorphic warnings are significantly more resistant to the habituation effect than conventional warnings. Despite the evident benefits, to the best of our knowledge, no warnings for phishing attacks in browsers or email clients implement this polymorphic behavior [22]. Only two recent studies combined the use of polymorphic behaviors with ad-hoc explanations to improve warning effectiveness [13, 21].

Another factor that can negatively affect the protection level of warnings is the *lack of specific explanations* about the risk related to the phishing attack [10]. Warnings should clearly state why a particular content is dangerous and what are the possible consequences of not complying with the alert. Instead, they often present generic messages that do not explain which are the elements of an email or a website that are dangerous to the user. This leaves the burden of making the final decision about suspicious content to the user, who is often not a cybersecurity expert and does not possess sufficient elements for making an informed decision [23]. On the other hand, including an explanation in the warnings can better help users understand the danger, addressing their lack of technical knowledge; explanations also have the potential to increase their motivation to follow the warning's suggested action and increase their trust in the system [10, 64]. [10, 64].

Warnings with explanations in the context of web browsers have been reported in [21]. The authors started by identifying a set of 7 website features that can be easily explained to users; then, they created two explanation messages for each feature. The generations of the explanations followed an iterative process that aimed at optimizing metrics of understandability, readability, and sentiment through manual tweaking of the explanation texts. The proposed warnings with explanations were more understandable and effective than the ones used by Google Chrome, Mozilla Firefox, and Microsoft Edge, which do not include explanations. A similar approach has been investigated in the context of email clients and explored both the active behavior and the explanation in warning dialogs [13]. The rationale of the proposed warnings was to prevent users from accessing visually deceptive phishing emails altogether, imitating what is done in modern browsers with malicious websites. The study showed evidence that combining active behavior and explanations was indeed more effective than state-of-the-art warnings for email clients proposed by Petelka et al. [54].

### 2.2 Importance of cyberpsychology in phishing attacks

Cyberpsychology is an interdisciplinary field at the intersection of psychology and technology and explores the relationship between humans and technology. Examples are the study of human experiences, including cognitive, emotional, and behavioral aspects influenced by technological advancements [55]. This discipline is also becoming more and more critical for cyber security as attacks are increasingly based on psychological aspects. Among the attacks that exploit such aspects there is the phishing attack, a scam that manipulates human behaviors and emotions to bypass security measures. Persuasive strategies [17] are often used by criminals to convince victims to open phishing links or download malware files. Phishers may, e.g., impersonate authoritative persons or institutions to scam victims (*authority* persuasion principle). A frequent example regards phishing emails sent from fake banks, government agencies, or corporate executives that might prompt users to divulge sensitive information without adequate scrutiny [32]. Phishers also exploit a sense of *urgency* or *scarcity* to force individuals to act quickly, bypassing critical thinking processes. This technique leverages the human tendency to prioritize immediate threats or opportunities over thorough analysis [68]. The principles of *reciprocity* and *social conformity* are also implemented by attackers in the email content; for example, phishing emails might promise a reward or request minor compliance in return for sensitive information, leveraging social norms to increase compliance [71].

Another psychological aspect phishers leverage is human emotions or emotional states, which might influence individuals' susceptibility to phishing. Phishing attacks frequently induce *fear* and *anxiety* to prompt quick, uncritical actions; in these cases, emails that threaten account suspension, legal action, or financial loss create a sense of urgency, compelling individuals to act quickly to avoid negative consequences. Vishwanath et al. proved that fear-based messages increase susceptibility to phishing attacks [65]. Another emotion exploited in this context is *curiosity*: emails that promise exclusive offers, sensational news, or intriguing content can trigger curiosity, leading individuals to click on malicious links or download harmful attachments. This was demonstrated by Goel et al. [30], who have shown that emails offering free gifts or sensational news increase phishing effectiveness. Phishers often craft messages that evoke *sympathy* by impersonating friends, family members, or trusted organizations: these messages might contain personal appeals for help, such as requests for financial assistance or urgent support. A study by Parsons et al. [53] found that messages appearing to come from known contacts were indeed significantly more successful.

While the previous aspects regard psychological aspects of the email content, further factors might impact phishing susceptibility. This is the case of the *cognitive load* caused by the user interface the user interacts with when reading the emails. High cognitive load resulting from complex or cluttered email clients or warnings can impair decision-making processes, making users more prone to be victims of phishing. Alsharnouby et al. indicate that simplifying the decision-making environment and providing clear, concise information can enhance users' ability to identify and avoid phishing attempts [3].

The integration of explanations into warnings has the potential to exercise a positive influence on user cognitive processes and behavior. For instance, explanations can prompt users to be more vigilant by providing specific reasons for suspicion. This may include highlighting inconsistencies in email addresses, grammatical errors, or requests for sensitive information, which can alert users to potential scams [24]. Furthermore, explanations are beneficial in enhancing the credibility and trustworthiness of warnings, increasing the probability that users will adhere to the warnings and take appropriate action [9]. Additionally, explanations are instrumental in facilitating users' cognitive processing. Providing clear explanations may assist users in comprehending the specific characteristics that make an email suspicious, thereby enhancing users' capacity to identify analogous threats in the future [25]. Also, explanations encourage users to engage in more analytical and less heuristic processing. By understanding the rationale behind an email being a scam, users are less likely to rely on superficial cues and more likely to evaluate the message content critically [70]. Explanations can also facilitate decision-making by identifying the most pertinent phishing indicators, thereby reducing the cognitive burden and enabling users to make well-informed security choices [59]. Finally, since explanations given in a warning change according to the specific email or website, another benefit of explanation is that including different messages in the warnings makes them behave in a polymorphic way, helping to lower the users' habituation to these messages [4].

### 2.3 LLMs in the context of phishing

Large Language Models (LLMs) are considered one of the most significant technological advancements in recent years, revolutionizing the field of Natural Language Processing. LLMs are based on the Transformer architecture [63] and are trained on massive datasets of text, potentially providing access to vast amounts of knowledge. The high performance of LLMs in various human tasks [35] is due to their large number of parameters, which enable them to identify intricate patterns in linguistic data. There are several commercial LLMs, with the most well-known being the OpenAI's GPT models, including GPT-4o[1]. Among other most performant LLMs, there are the latest Gemini models (1.0 Ultra and Gemini 1.5 Pro) by Google[2], Claude3.5 Sonnet by Anthropic[3], and Meta's Llama 3[4].

Interaction with LLMs usually happens in natural language using messages called "prompts". The discipline of prompt engineering has recently emerged to guide the design of prompts and optimize them to use LLMs at their best potential, obtain the desired inputs, and understand their limitations [20]. There are several guidelines for writing a proper prompt [20, 60], which are generally easy to apply but can still drastically improve the quality of the results. A famous technique is called "few-shot prompting" [18] and consists of providing the LLM with examples of input-output interactions to define the structure and style of the desired output. Few-shot prompting is somehow affine to fine-tuning, in the sense that the model receives a small number of domain-specific examples to adjust its outputs, but the former is often preferred over the latter [45, 58].

Recently, LLMs have been used as cybersecurity tools: an example of such an application is the analysis of malicious scripts to explain the potential danger to cybersecurity analysts [28]. Another security application is in the context of phishing; LLMs can also be used to process and classify textual information like emails and websites and detect phishing content. GPT-2 has been shown to be promising in the task of detecting phishing emails in earlier studies [48], including adversarial settings [29]. The technological advance in LLMs seems to lead to substantial improvement in the phishing detection task. For example, in [41] Koide et al. proposed a method to detect phishing websites with LLMs, comparing GPT-3.5 and GPT-4; the latter model achieved substantially better results compared to GPT-3.5, achieving a 98.4% accuracy (vs. the 92.6% accuracy of GPT-3.5), with a significant improvement in GPT-4 in avoiding false negatives. In the work of Heiding et al. [33], four LLMs (GPT-4, Claude-1, Bard, and LLaMA2) were used to detect the intention (malicious vs. genuine) of phishing emails generated by humans and by GPT-4. All the LLMs proved to be able to detect malicious content, even in non-obvious phishing emails, effectively. Results also suggest that LLMs might be better at detecting phishing emails compared to humans. An interesting aspect of this work relies on the results obtained when priming LLMs to look specifically for malicious intent in the emails: a simple change in the prompt (asking it to look for "malicious" intent rather than asking for generic intent) led to a substantial increase in detection performance, without affecting the false-positive rate.

---

[1] https://openai.com/index/hello-gpt-4o
[2] https://deepmind.google/technologies/gemini
[3] https://www.anthropic.com/news/claude-3-5-sonnet
[4] https://llama.meta.com/llama3

## 3   APOLLO: AN LLM-POWERED SYSTEM FOR DEFENCE AGAINST PHISHING ATTACKS

The first contribution of this study is *APOLLO*, a tool written in Python 3.10.12 and powered by GPT-4o to (i) classify an email as phishing or legitimate and (ii) generate an explanation for the user in case of a phishing email. As an LLM model, we chose to use GPT-4o because, at the time of writing this article, it resulted to be the best-performing model for classification-like tasks[2], such as the one performed by our system. The APOLLO source code can be found at this link. APOLLO is composed of 3 main modules (see Figure 1), which are detailed in the next subsections: the *preprocessor* module, the *URL enricher* module, and the *LLM prompter* module.

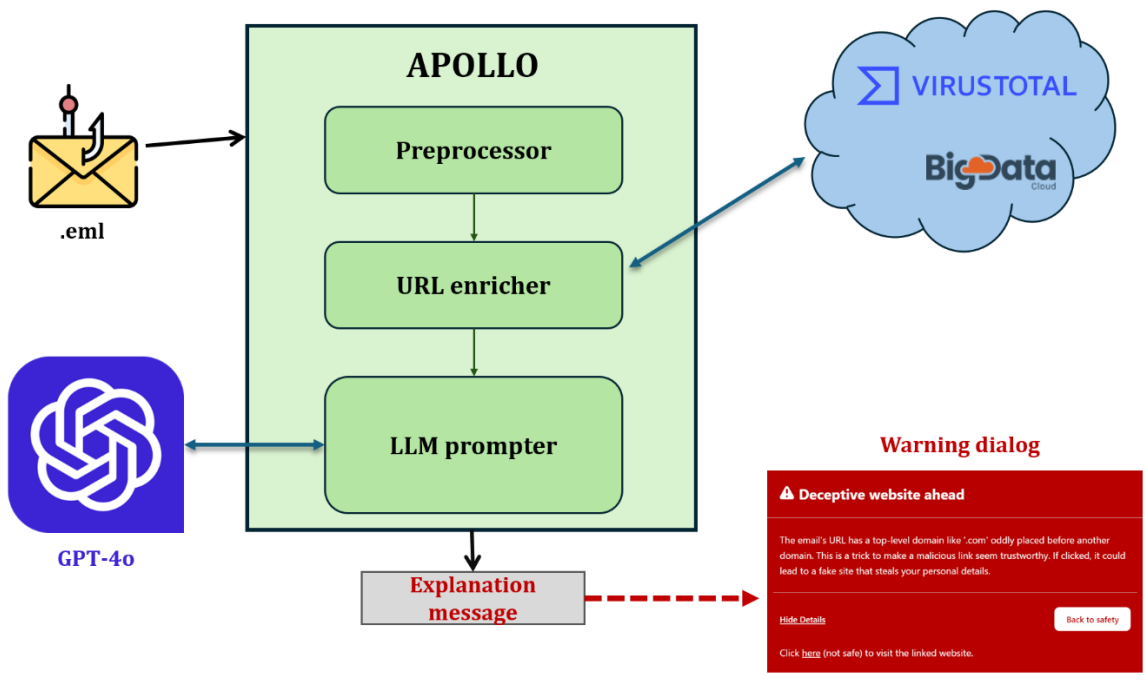

Figure 1. The architecture and information flow of APOLLO

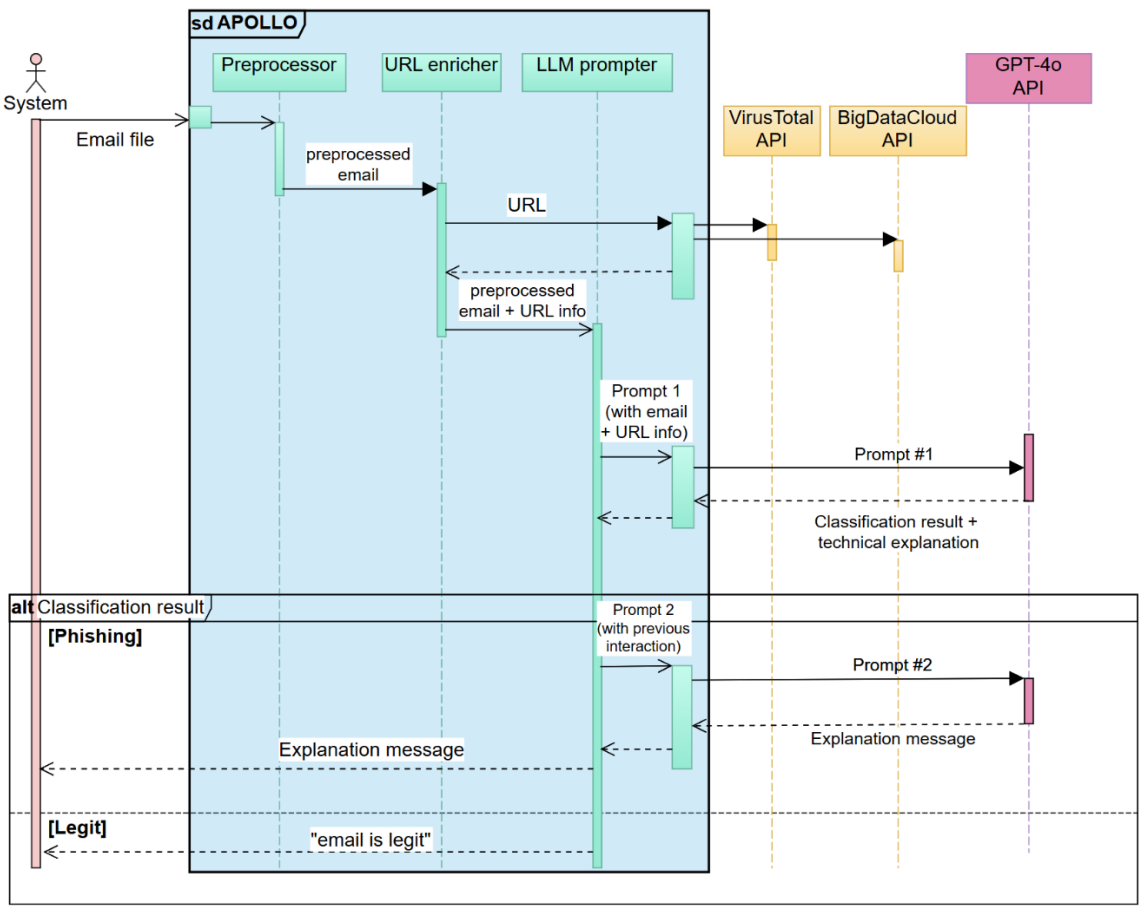

Figure 2. Sequence diagram of APOLLO. The tool takes an email file as input and returns the explanation message to insert in a warning dialog if the email resulted to be phishing according to the GPT model.

## 3.1 Preprocessor module

The first module in the pipeline of APOLLO is the *preprocessor* module, which takes an email in .eml format in input and makes it more easily readable for the GPT model. This is done by following a similar approach to the preprocessing applied in the work of Misra and Rayz [48]:

1. The subject of the email is extracted; if no subject is found, then it is set to "NO SUBJECT".
2. The email headers are extracted and inserted in a dictionary (named array).
3. The email body is extracted, and all HTML tags (if any) are removed by using the `BeautifulSoup` Python library[5].
4. Information about any URL, phone number, and email address in the email body is saved by applying a special meta-tag in place of the HTML anchor tag; specifically, meta-tags "URL", "PHONE", and "EMAIL" are used, respectively.
5. Other preprocessing is performed to remove subsequent blanks or newline characters.

---
[5] https://beautiful-soup-4.readthedocs.io/en/latest/

The preprocessing procedure is required for different goals. First, it reduces the length of the text to feed into the GPT model; this is essential since the context window is limited and might not be sufficient for treating a raw .eml file containing several headers, an HTML body, etc.; this is true especially if one decides to use a smaller GPT model. Another advantage of applying a preprocessing step is that of giving the LLM less textual information to work on, which could improve results [15].

At the end of the preprocessing step, the original email is broken down into: the dictionary of headers, the subject, the preprocessed body, and the list of URLs found in the email. The URLs will then be fed to the URL enricher module to extract additional information.

### 3.2 URL enricher module

The *URL enricher* module is needed to retrieve threat intelligence to give the GPT model more grounded facts on which to reason; notably, this is also done to overcome the knowledge cut-off of the adopted LLM model (e.g., *gpt-4-1106-preview*, which was used in our user study, is trained on data up to April 2023 [51]). This approach is a form of Retrieval Augmented Generation [43], as it gathers information from external sources to improve the accuracy and reliability of the LLM model. In this version of APOLLO, we do not consider every URL retrieved in an email, but only the first one (in order of appearance). This decision brings some limitations to the tool but simplifies its development considerably. Therefore, from the first URL, the tool extracts the full hostname ("protocol://hostname", e.g. "https://www.google.com"), which we will call just URL for simplicity, and feeds it to 2 API services:

1. VirusTotal[6], a very popular security tool for scanning files and URLs; feeding a URL into the API endpoint https://www.virustotal.com/api/v3/urls/{base_64_encoded_url} allows scanning it with more than 90 antivirus products and blocklists to produce a threat score for each of them. From the results of a scan, APOLLO takes the number of votes for the URL, which can be either *harmless*, *undetected*, or *malicious*.
2. BigDataCloud[7], an IP geolocation service to retrieve the server location of the website linked by a URL. From the URL, the IP address is extracted with Python's `dns.resolver` module, and then it is used to call the endpoint https://api.bigdatacloud.net/data/country-by-ip?ip={ip_addr}. From the retrieved results, the tool takes just the ISO 3166-1 alpha-3 country code (e.g., "ESP" for Spain), which is enough for GPT to identify a country univocally.

All these services are free to use under the constraint of a low API call rate. At the end of the URL enrichment step, the first URL will be correlated with the VirusTotal scans list and the country code of the linked domain.

### 3.3 LLM prompter module

The *LLM prompter* module finally takes the information obtained from the two previous steps and feeds them into the GPT model to produce a classification outcome for the email and an explanation. This is done by sequentially filling two prompts with the email data. These prompts constitute the core of APOLLO. Therefore, the design of the prompts required considerable manual effort and several iterations to refine them according to best practices of prompt engineering and empirical observation of the outcomes for different inputs [20, 45, 60]. The prompts were also iteratively tested on a small subset of phishing and legitimate emails in the inbox of one of the authors; the outcomes were evaluated based on (i) whether the classification outcome was right or wrong, (ii) whether the phishing cues and the social engineering techniques indicated by GPT were meaningful (we did not want to force the model in producing hallucinated outputs), and (iii) whether we considered the generated explanations easy enough for non-experts to understand.

---

[6] https://www.virustotal.com
[7] https://www.bigdatacloud.com/ip-geolocation

---

First prompt. This makes GPT generate a classification outcome for an email and generate a first explanation

---

You are a cybersecurity and human-computer interaction expert who has the goal of detecting if an email is legitimate or phishing and helping the user understand why a specific email is dangerous (or genuine) in order to make more informed decisions.

The user will submit the email (headers + subject + body) optionally accompanied by information on the URLs in the email as:

- server location;

- VirusTotal scans reporting the number of scanners that detected the URL as harmless, undetected, malicious.

Your goal is to output a JSON object containing:

- The classification result (label).

- The probability as a percentage of the email being phishing (0%=email is surely legitimate, 100%=email is surely phishing) (phishing_probability).

- A list of persuasion principles that were applied by the alleged attacker (if any); each persuasion principle should be an object containing:

   the persuasion principle name (authority, scarcity, etc.),

   the part of the email that makes you say that the persuasion principle is being applied;

   a brief rationale for each principle.

- A list of 3 to 5 features that could indicate the danger (or legitimacy) of the email; the explanations must be understandable by users with no cybersecurity or computer expertise.

Desired format:

label: <phishing/legit>

phishing_probability: <0-100%>

persuasion_principles: [array of persuasion principles, each having: (name, specific sentences, rationale)]

explanation: [array of 3-5 features explained]

Email:
[HEADERS] *{email headers}* [\HEADERS]
[SUBJECT] *{email subject}* [\SUBJECT]
[BODY]
*{email body}*
[\BODY]
URL Information:
Server location: *{URL geolocation}*
VirusTotal scan: [
	harmless: *{n_harmless}*,
	undetected: *{n_undetected}*,
	malicious: *{n_malicious}*
]

The first prompt (reported in the following) is needed to initialize the session and make GPT produce a JSON object that contains the result of the classification (phishing/legitimate and a probability), a list of possible persuasion principles that were applied in the email, and a list of phishing (or legitimacy) indicators, each with an explanation. This list ranges from 3 to 5 items by default. This prompt is followed by a message containing the preprocessed email and the URL information. The email itself is divided into its headers, subject, and body, while the URL information consists of the geolocation information, and VirusTotal data (i.e., the number of detectors that classified the URL as "harmless", "undetected", and "malicious").

Starting from the data obtained as a result of the LLM given the first prompt, a more refined explanation is generated using a second prompt (reported in the following). This multiple-step approach is referred to as "Prompt Chaining" [19] and is useful when we need to generate responses that must undergo different operations or transformations to improve the output quality. We wanted the generated explanations to follow the structure presented in [21] since it is already well-defined and grounded in warning theory for the design of warning messages [6], i.e., "Feature description + Hazard Explanation + Consequences of not complying with the warning".

To make GPT generates explanations that follow this structure, we opted for a *few-shot prompting* approach, as also suggested by OpenAI [60], by also supplying three examples of desired explanation messages. These examples were taken from the work of Desolda et al. [21], since they were manually and carefully designed according to different metrics of readability, understandability, and sentiment.

---

Second prompt. This generates a refined explanation message following the desired structure (as in [21])

Now take the most relevant feature among the ones in your explanations and construct a brief explanation message (max 50 words) directed to naive users (with no knowledge of cybersecurity) that will follow this structure:

1. description of the most relevant phishing feature
2. explanation of the hazard
3. consequences of a successful phishing attack

For example, an explanation that explains that the top-level domain in one of the email's URLs is mispositioned would be:

"In the URL present in the email the top-level domain is in an abnormal position. This could indicate that the URL leads to a fake website. Such websites might steal your personal information".

Another example of an explanation about the domain of a website being suspiciously young would be:

"The URL in the email leads to a website created N days ago. Young websites are famous for criminal activity. There is a potential risk if you proceed."

Another example of explaining that the email is suspicious because of too many special characters in its body would be:

"Many special characters have been detected in the email. Malicious people use them to disguise text and deceive you. Your data could be stolen."

Desired format:

[description of the feature]. [hazard explanation]. [consequences of a successful attack].

---

To have more control over the explanations generated by APOLLO, we can make it generate explanations that regard a specific phishing feature (e.g., that a URL in the email leads to a very young web domain) by "priming" the GPT model. To force the generation of a feature-specific explanation, we can change the prompt by including: (i) the name of the feature to explain and (ii) a description of the feature. We have to consider that, by priming GPT in this way, we might make it hallucinate, as it will try to generate a meaningful text, even if not grounded on true facts. This feature was, nonetheless, fundamental for our study, as we wanted to include explanations only for email features that were already studied in the literature and that resulted in being valuable for users in making decisions regarding phishing content [13, 21], i.e.:

1. The Top-Level Domain of a URL in the email is mispositioned (e.g., as in the URL "www.amazon.com.cz").
2. A URL in the email is an IP address instead of a normal hostname.
3. The mismatch between a link shown in the email and its actual destination in the email (e.g., "click here" is shown instead of the URL itself).
4. A URL in the email points to a very young domain (as phishing websites are generally hosted on newly created domains).

To generate explanations for specific features, we simply modified the starting sentence of the second prompt as "Consider that the previous email is suspicious because…" followed by a brief description of the feature.

## 4 EVALUATING THE PERFORMANCE OF APOLLO

### 4.1 Materials and Methods

To determine the performance of APOLLO in detecting phishing emails, we conducted an evaluation of the system in terms of accuracy in discerning between phishing and genuine emails. We tested the system by using a dataset of 4000 emails (half phishing, half genuine) sampled from the Nazario, NigerianFraud, and SpamAssassin datasets [16]. The sampling criteria were 1) emails with at least one link and ii) the most recent emails in the datasets. The final dataset used in our evaluation is available in the APOLLO repository. As an LLM, we used GPT-4o (specifically, version *gpt-4o-2024-05-13*), the most recent and performing OpenAI model for classification tasks when writing this paper [52].

A Python script (*evaluation.py*) was created to perform the evaluation. The assessment starts by accessing the dataset and invoking the preprocessor and LLM prompter modules in sequence for each email. The preprocessor module was updated given that the emails in the dataset were not in .eml files but as tuples <body, sender, receiver, date, subject, URLs, label>. From the answer of GPT-4o to the first prompt (the second one was not invoked since the explanation is not necessary to assess the classification performance), the "label" and "probability" fields were extracted: the former represents the crisp label ("phishing"/"legitimate") predicted by the GPT-4o model, while the latter is the estimated probability of the email belonging to the "phishing" class, ranging from 0 (surely legitimate) to 1 (surely phishing). These values were saved to the predicted_labels.xls file (in the "Classification evaluation/results/predicted labels" folder).

A significant element of this assessment is the evaluation of the potential benefits that the information on URLs provided by VirusTotal could offer to enhance classification performance. In fact, most emails used for evaluation date back more than ten years, and VirusTotal now almost always successfully labels the old URLs. This behavior, however, does not reflect what the classification of the same emails by VirusTotal would have been over time. In fact, these services often do not immediately classify the emails or related links as malicious, and this happens because new phishing attacks may not yet be immediately recognized. Typically, their classification may become stable within a few hours or days, depending on the speed with which new samples are analyzed and antivirus signatures updated [5, 62]. Therefore, if we evaluated our system using today's values of VirusTotal, we would not be able to understand the extent to which VirusTotal information, which changes over time, impacts classification.

Therefore, we simulated an evaluation considering the trend at different accuracy rates of these services. To do this, we first determined the values that the parameters of VirusTotal can take on real genuine and phishing URLs. These ranges were determined by invoking the VirusTotal API on the URLs of the 4000 emails stored in the dataset used for the evaluation; in addition, to improve the external validity of the results, we also performed the same analysis on a further 4000 URLs sampled from the URLs stored in the PhishTank repository[8] (the details of this analysis are reported in the project repository, folder *Classification evaluation/results/VirusTotal ranges*). We empirically observed that, although VirusTotal claims to use over 90 detectors, in the case of old or well-known malicious URLs, only a small subset of them succeeds in detecting malicious URLs. In contrast, other detectors work mainly on detecting malicious files. The resulting parameter ranges are: $n\_harmless$ = [0-87], $n\_undetected$ = [0-28], $n\_malicious$ = [0-25]. These ranges show, for example, that the maximum number of VirusTotal detectors for malicious URLs is 25.

The evaluation started by determining the model's performance considering no information about the URL; the aim was to assess the raw performance of GPT-4o on the email phishing detection task as is. Then, we ran the evaluation using different versions of URL information in the prompts to simulate different accuracy rates in the output received by the URL enricher module. To do this, we simulated the output returned by the VirusTotal API at five "percentiles" $Q$ of the three parameters. We started by modeling the best-case scenario ($Q100$), i.e., the case in which the output from VirusTotal strongly confirms the legitimacy/maliciousness of a URL. Then, we progressively decreased the confidence of this information ($Q75, Q50, Q25$), arriving at the case in which the information from VirusTotal is entirely uncertain, and the URL is considered neither malicious nor harmless ($Q0$), a typical case of data that is returned by VirusTotal at time zero of an attack; in this case, both genuine and phishing URLs are reported as "undetected" by 100% of VirusTotal's detectors. The precise information about the simulated values used in the GPT-4o prompt for the evaluation is reported in Table 1.

Using a similar rationale, we simulated wrong information coming from VirusTotal, i.e., how different degrees of errors in the URL information can affect the final classification of GPT-4o. Even if this is a rare situation, given the high quality of this service, it is interesting to analyze how GPT-4o classifies emails when the prompt contains misleading information. Therefore, we started by introducing a small error, such as phishing emails being detected as "harmless" by 25% of detectors (in the previously specified range) and legitimate emails being detected as "malicious" by 25% of detectors (Q25ERR); the error was increased to 50%, 75% and 100% for the remaining conditions, respectively Q50ERR, Q75ERR, Q100ERR. The specific values used for the percentiles can be found in Table 1.

Table 1. Simulated VirusTotal data for each condition. We simulated the number of detectors reporting the URL as "harmless", "undetected", or "malicious". The two sub-columns under each column report the number of harmless, undetected, or malicious detectors in the cases of emails classified by GPT-4o as genuine (in blue) and phishing (in red).

| Conditions | n_harmless | | n_undetected | | n_malicious | |
|---|---|---|---|---|---|---|
| | legit | phishing | legit | phishing | legit | phishing |
| noURL | N/A | N/A | N/A | N/A | N/A | N/A |
| Q0 | 0 | 0 | 28 | 0 | 0 | 0 |
| Q0 | 0 | 0 | 28 | 0 | 0 | 0 |
| Q25 | 22 | 0 | 21 | 0 | 0 | 6 |
| Q50 | 43 | 0 | 14 | 0 | 0 | 12 |
| Q75 | 65 | 0 | 7 | 0 | 0 | 19 |
| Q100 | 87 | 0 | 0 | 0 | 0 | 25 |
| Q25ERR | 0 | 22 | 21 | 0 | 6 | 0 |
| Q50ERR | 0 | 43 | 14 | 0 | 12 | 0 |
| Q75ERR | 0 | 65 | 7 | 0 | 19 | 0 |
| Q100ERR | 0 | 87 | 0 | 0 | 25 | 0 |

---

[8] https://phishtank.org/developer_info.php

As for the geolocation information, the external service of BigDataCloud was used within every evaluation that included the URL information. Therefore, in every condition aside from the *noURL* one, the emails were accompanied by the actual country in which the linked web domain was located. It is worth noting that not isolating the individual effects of BigDataCloud and VirusTotal on the classification outcomes represents a limitation of this study that will be addressed in future evaluations.

To limit the fluctuation in the results of the GPT model and obtain more deterministic outputs, we set the TEMPERATURE parameter in the LLM prompter module to a very low value (0.0001) throughout the evaluation process [50]. Moreover, to test if the outputs of different evaluations remained stable over different executions, we repeated the evaluation of the GPT-4o model without the URL information 5 times.

### 4.2 Data analysis

Precision, recall, accuracy, and F1-score are used to measure the model's performance in the email binary classification task. Moreover, log-loss and ROC AUC were used to measure the performance in the probability estimate for the two classes. Precision measures the proportion of true positive predictions among all positive predictions made by the model. Recall measures the proportion of true positive predictions among all actual positive cases. Accuracy measures the proportion of correct predictions (both true positives and true negatives) among all predictions made by the model. The F1-Score is the harmonic mean of precision and recall, balancing the two metrics. Log-loss measures the performance of a classification model where the prediction is a probability value between 0 and 1; it is the negative log-likelihood of the true labels given the predicted probabilities, and a lower log-loss indicates better performance. The ROC AUC measures the area under the ROC curve, which plots the true positive rate (recall) against the false positive rate (1-specificity) at various threshold settings; AUC represents the probability that a randomly chosen positive instance is ranked higher than a randomly chosen negative one. An AUC of 1 indicates a perfect model, while an AUC of 0.5 suggests a model no better than random guessing.

The chi-square test was employed as an omnibus test to verify any differences in the predicted labels (dichotomous nominal value) among various conditions. Subsequently, the chi-square test was employed as a post-hoc test with Bonferroni correction in case of significant differences. Similarly, the One-way ANOVA was performed to assess any differences in the predicted probability; Tukey's HSD post-hoc tests were performed in case of significant differences. All the metrics details, predicted labels, predicted probabilities, and inferential tests are reported in the project repository under the folder "*results/Classification evaluation*".

### 4.3 Experimental Results

The results of the evaluation show that the performance of GPT-4o in the email phishing detection task is very high, even without including additional information about the URL. Indeed, it resulted in a precision of 0.964, a recall of 0.985, and an accuracy of 0.974, indicating that the model can effectively distinguish between phishing and legitimate emails based solely on the email content and metadata; moreover, the low log-loss (0.113) and high ROC AUC (0.994) further confirm that the model provides reliable probability estimates and maintains excellent discriminative power (see Table 2).

A chi-square test of independence was performed to examine the relation between the different conditions in the cases where the right URL information is present or not. The relation between the examined conditions (noURL, Q0, Q25, Q50, Q75, Q100) and the predicted label was significant ($X^5 = 653.39$, $p < .001$). Thus, post-hoc tests were performed among all the pairs; the full data is reported in Appendix A, while the resulting p-values are reported in Table 3. The results of the pairwise comparisons indicated that by including the information on the URL when the attack starts (Q0), the accuracy decreases with respect to noURL ($p = <.001$). This result shows how insignificant information returned from a third party service (e.g. n_harmless=0, n_malicious=0) can confuse the classification of the model, making it worse than it would be without such information; we can conclude that in case of such information on the URL, the parameters and their values should be excluded from the prompt. The benefits of URL information begin to become apparent in the next quartiles. Indeed, in the case of Q25, the model's performance is almost perfect (F1=0.999), outperforming both the noURL ($p = <.001$) and Q0 ($p = <.001$) conditions. The same level of performance was observed in the case of Q50, Q75, and Q100, as in these instances, the model incorrectly classified only four emails.

Table 2. Performances of APOLLO in the binary classification task. The performance of GPT-4o becomes increasingly precise as the information about the URL becomes more certain (Q≥25) to a maximum at Q100, while adding uncertain information (Q0) decreases performance.

| Conditions | Prediction | | Predicted classes metrics | | | | Predicted probabilities metrics | |
|---|---|---|---|---|---|---|---|---|
| | Correct | Wrong | precision | recall | accuracy | F1-score | log-loss | ROC AUC |
| noURL | 3896 | 104 | 0.964 | 0.985 | 0.974 | 0.974 | 0.113 | 0.994 |
| Q0 | 3797 | 203 | 0.997 | 0.901 | 0.949 | 0.947 | 0.279 | 0.981 |
| Q25 | 3996 | 4 | 0.998 | 1.000 | 0.999 | 0.999 | 0.804 | 0.962 |
| Q50 | 3996 | 4 | 0.998 | 1.000 | 0.999 | 0.999 | 0.753 | 0.968 |
| Q75 | 3996 | 4 | 0.998 | 1.000 | 0.999 | 0.999 | 0.254 | 0.992 |
| Q100 | 3996 | 4 | 0.998 | 1.000 | 0.999 | 0.999 | 0.136 | 0.997 |
| Q25ERR | 1795 | 2205 | 0.473 | 0.898 | 0.449 | 0.619 | 1.663 | 0.356 |
| Q50ERR | 1775 | 2225 | 0.470 | 0.888 | 0.444 | 0.615 | 1.911 | 0.336 |
| Q75ERR | 1790 | 2210 | 0.472 | 0.895 | 0.448 | 0.618 | 4.686 | 0.267 |
| Q100ERR | 1798 | 2202 | 0.473 | 0.899 | 0.450 | 0.620 | 15.230 | 0.027 |

Despite the precision, accuracy, recall, and F1-score are always the same when Q≥25, it is noteworthy that the overall performance of GPT-4o becomes increasingly precise as the information about the URL becomes more certain to a maximum at Q100, which led to excellent metrics in log-loss (0.136) and an almost perfect ROC AUC (0.997). These results demonstrate that integrating reliable external data might enhance the model's performance. Tukey's HSD post-hoc tests applied to the predicted probabilities indicated that only the Q25 condition led to superior outcomes compared to Q0, while no differences were observed in the remaining comparisons. This result shows that very strong performances are already achieved at the 25th percentile, while improvements are not significant in the subsequent percentiles.

Further interesting outcomes come from the evaluations of GPT-4o in the case of wrong information from VirusTotal. Indeed, the results (see Table 2 and Table 4) indicate that even a small error (Q25ERR) of these services may seriously hinder the model's performance significantly. Increasing the amount of error in the prompt (Q50ERR, Q75ERR, Q100ERR) increases the error in the predicted class probabilities (i.e., the log loss), making the model increasingly confident in giving a wrong outcome. However, injecting wrong URL information mostly affects GPT-4o's precision rather than its recall; this means that introducing this type of error can easily lead to false positives (i.e., wrongly alerting about genuine emails) rather than false negatives (i.e., not alerting about phishing emails). These results suggest that GPT-4o adopts a cautious approach by default and is robust to the introduction of wrong information, detecting phishing emails accurately in any case.

Table 3. P-values resulting from the post-hoc tests performed with the chi-square test (lower triangular matrix – in green) and Tukey's HSD test (upper triangular matrix – in red) for the *Q0-Q100* conditions. No significant gain in performance is observed for values above the 25th percentile.

|       | noURL | Q0   | Q25  | Q50  | Q75  | Q100 |
|-------|-------|------|------|------|------|------|
| noURL |       | .000 | .353 | .283 | .811 | .876 |
| Q0    | .000  |      | .000 | .000 | .000 | .000 |
| Q25   | .000  | .000 |      | .000 | .000 | .000 |
| Q50   | .000  | .000 | 1    |      | .000 | .000 |
| Q75   | .000  | .000 | 1    | 1    |      | .000 |
| Q100  | .000  | .000 | 1    | 1    | 1    |      |

Table 4. The p-values resulting from the post-hoc tests performed with the chi-square test (lower triangular matrix – in green) and Tukey's HSD test (upper triangular matrix – in red) for the *Q25ERR-Q100ERR* conditions, considering *noURL* as a baseline. A significant worsening in performance is observed for values above the 25th percentile when prompts are augmented with erroneous information, compared to not including any information about URLs (*noURL*).

|          | noURL | Q25ERR | Q50ERR | Q75ERR | Q100ERR |
|----------|-------|--------|--------|--------|---------|
| noURL    |       | .000   | .000   | .000   | .000    |
| Q25ERR   | .000  |        | .732   | .845   | .000    |
| Q50ERR   | .000  | 1      |        | .243   | .000    |
| Q75ERR   | .000  | 1      | 1      |        | .000    |
| Q100ERR  | .000  | 1      | 1      | 1      |         |

The last evaluation regards the assessment of the fluctuation of GPT-4o. To this aim, we repeated the evaluation of the *noURL* condition a total of 5 times, obtaining very similar results (reported in Table 5). The chi-square test did not lead to any significant difference in the classification outputs ($\chi(1) = 1.936$, $p = .747$). On the contrary, there was a statistically significant difference between the predicted probabilities as determined by the one-way ANOVA ($F(4,19873) = 16.016$, $p < .001$). A Tukey post-hoc test revealed that this difference is caused by the probabilities of the third repetition, which resulted statistically lower than the other repetitions (detailed results in the repository under the folder "results/Classification evaluation"). These findings suggest that while GPT-4o's classification decisions remain stable, the confidence (as represented by predicted probabilities) in these decisions can vary, indicating potential sensitivity to certain conditions or inherent model variability. Further research should investigate the underlying reasons for the significant differences in predicted probabilities, particularly focusing on the conditions or model states during the third repetition. Moreover, solutions to enhance the stability of the predicted probabilities without compromising the classification accuracy should be investigated.

The triangulation of all the previous results allows us to determine the optimal conditions for using GPT-4o for the identification of phishing emails. The classification of an email must start with an examination of the external service data, which must be sufficiently mature (Q>25 in the case of VirusTotal) to be employed profitably within the prompt; otherwise, it would be optimal to discard external data, as highly uncertain information (Q0) showed to hinder the performance of the model. Then, the LLM can be invoked to obtain the classification (phishing/genuine) and probability. The combination of LLM with the third-party service resulted in a near-perfect accuracy (99%), which is higher than the 98.4% accuracy obtained by Koide et al. [41] using only GPT-4. Furthermore, the log-loss information on the classified email can be used to decide whether the model is confident that the email is correctly classified (e.g. log-loss<1, considering the results reported in Table 2) or not (log-loss>=1). This information can be used to generate different explanations in the warnings, i.e. to communicate the uncertainty of the prediction to the users or not.

Table 5. Repeated evaluation metrics for the noURL condition. A total of 5 repetitions were performed.

| # Repetition | Prediction | | Predicted classes metrics | | | | Predicted probabilities metrics | |
|---|---|---|---|---|---|---|---|---|
| | **Correct** | **Wrong** | *precision* | *recall* | *accuracy* | *F1-score* | *log-loss* | *ROC AUC* |
| *1* | 3896 | 104 | 0.964 | 0.985 | 0.974 | 0.974 | 0.113 | 0.994 |
| *2* | 3895 | 105 | 0.962 | 0.985 | 0.973 | 0.973 | 0.114 | 0.994 |
| *3* | 3909 | 91 | 0.965 | 0.986 | 0.977 | 0.975 | 0.108 | 0.995 |
| *4* | 3890 | 110 | 0.962 | 0.984 | 0.973 | 0.973 | 0.133 | 0.993 |
| *5* | 3897 | 103 | 0.963 | 0.986 | 0.974 | 0.975 | 0.103 | 0.995 |

## 5 EVALUATING THE WARNINGS GENERATED BY APOLLO

In order to evaluate the qualities of warnings generated by APOLLO, we conducted an explorative user study to measure how users perceived these warnings in the context of an email client. This section presents the results of this study, plus a comparison of these results to state-of-the-art solutions. Four warnings were created using APOLLO; as a baseline we used the study of Desolda et al. [21], considering four state-of-the-art warnings. To compare our warnings with the baselines in the most rigorous way possible, we replicated the study reported in the paper [21] in terms of design, user recruitment, scenario, and metrics.

### 5.1 Research questions, study design, and participants

The aim of this study was to answer the following research question: "*How do users perceive warning dialogs for phishing attacks generated by LLMs?*". We adopted a within-subjects design with the *warning* being the independent variable, and the 8 within-subject levels being the 4 warnings proposed in this study, plus the 4 warnings of [21] as a baseline: the first baseline warning includes explanations created manually (this is called *manual explanation* in the following), while the other three are respectively the warnings of Google Chrome, Mozilla Firefox and Microsoft Edge, which do not provide any explanation. The baselines were chosen because the *manual explanation* proved to be a valuable and effective warning that includes explanations, similar to the warnings in our study, and because the other warnings are those commonly used by most users [27].

Our study consisted of an online survey hosted on the *Lime Survey* platform. Participants were exposed to 4 phishing emails, each supplied with a specific warning. To recruit participants, we used the online platform Prolific (https://www.prolific.co); 20 participants took part (9 males and 11 females, average age of 30.85, sd=9.57).

### 5.2 Instruments and Measurements

We used APOLLO to generate explanation messages for 4 phishing emails that were heavily inspired by real emails; one of these emails was a genuine email (from Facebook) and was used as a false positive in our study. APOLLO recognized the email as genuine; therefore, we had to force it to generate a (wrong) explanation based on a feature that was observable in the email but that was harmless. The generated explanations were then used to build 4 warnings: we replicated the warning dialog of the *manual explanation* baseline and replaced the explanation message with the explanations generated by APOLLO. An example is reported in Figure 3, where the explanation is the sentence in the middle section ("The email's URL has a..."). We used the existing warning interface to make our explanations comparable to those in the baseline without introducing any other element in the visual interface. The 4 warning messages generated by using APOLLO are reported in Table 6. The original emails can be found at this link and are reported in Appendix B. It is noteworthy that at the time of the study's execution (January 2024), the most advanced LLM model available was GPT-4 turbo, which was employed for explanation generation. However, the final version of APOLLO released with this research integrates the more recent GPT4-o model, as it enhances performance on the classification task.

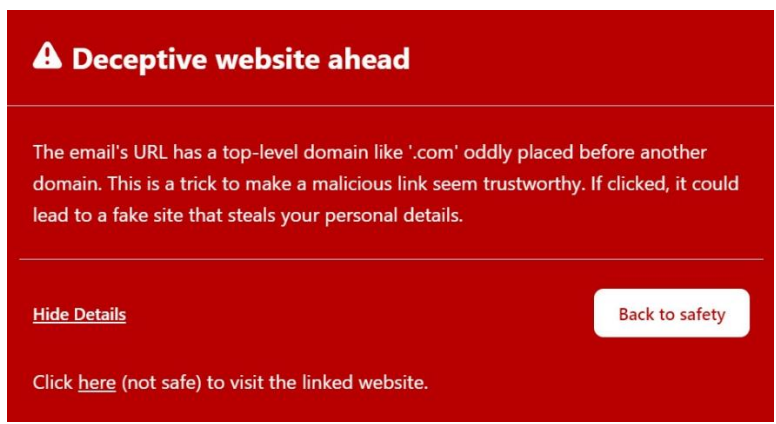

Figure 3. Example of warning dialog with an explanation message (W2) generated with APOLLO.

Table 6. Explanation messages for each condition.

| Warning | Explanation Message | Explained feature |
|---|---|---|
| **W1** | *The email contains a link that is just a string of numbers (an IP address). Legitimate companies usually use a name, not numbers. Clicking on it could lead to a fake site that steals your information.* | URL is an IP address |
| **W2** | *The email's URL has a top-level domain like '.com' oddly placed before another domain. This is a trick to make a malicious link seem trustworthy. If clicked, it could lead to a fake site that steals your personal details.* | Top-Level domain in the URL is mispositioned |
| **W3** | *The email shows a link labeled "protect your account", but it points to a different, suspicious website. This mismatch can trick you into visiting a harmful website. You might unknowingly give away your personal information or passwords* | Shown and actual links mismatch |
| **W4** | *The URL in the email leads to a very new domain. New domains are often used by scammers for fraud. You could be tricked into giving away personal details or downloading harmful software.* | Newly created domain |

To answer the research question, we collected quantitative and qualitative data through two questionnaires. The first one is the same questionnaire proposed in the study of Desolda et al. [21] and aims to measure different aspects of a warning dialog (we will call it "*warning questionnaire*" for brevity). The items of this questionnaire and the possible answers for each question are reported in Table 7. This questionnaire measures different aspects of warning dialogs for phishing attacks by considering dimensions such as understandability, familiarity, interest, perceived risk, and level of trust through Likert scales (quantitative data); additionally, it also measures aspects like the reaction, confusing words, perceived meaning and action to take through open-ended questions (qualitative data). Items 1 and 11 were control questions used to ensure that the participants read the warning and paid attention in order to detect and remove any inattentive participants, who are typically found in unsupervised remote studies [47].

Table 7. The questionnaire used to evaluate the warnings (*warning questionnaire*)

| # | Questionnaire item | Possible Answers |
|---|---|---|
| 1 | Did you read the entire text of the warning dialog? | [yes; partially; no] |
| 2 | When you saw the warning dialog, what was your first reaction? | [free text] |
| 3 | I understood the warning dialog | [5-point Likert scale, from "Strongly disagree" to "Strongly agree"] |
| 4 | I am familiar with this warning dialog | [5-point Likert scale, from "Strongly disagree" to "Strongly agree"] |
| 5 | I am not interested in this warning dialog | [5-point Likert scale, from "Strongly disagree" to "Strongly agree"] |
| 6 | Which word(s) did you find confusing or too technical? | [free text] |
| 7 | Please rate the extent of risk you feel you were warned about | [very low risk; low risk; no risk; risky; very high risk] |
| 8 | What action, if any, did the warning dialog want you to take? | [to continue to the website; to be careful while continuing to the website; to not continue to the website; I did not feel anything] |
| 9 | What do you think this warning dialog means? | [free text] |
| 10 | Please rate your level of trust in this warning dialog | [not at all confident; not very confident; neutral; confident; very confident] |
| 11 | What is the first word in this warning dialog? | [free text] |

The second one is the System Causality Scale (SCS) [34], a questionnaire that measures the perceived quality of explanations quantitatively. For each warning to which participants were exposed, they had to fill in both questionnaires, resulting in a total of 8 (2 questionnaires x 4 conditions) questionnaires answered by each participant. The items of the SCS are statements for which the participants had to express their agreement (or disagreement) on a 5-point Likert scale – from "Strongly disagree" to "Strongly agree". The complete list of items of the SCS questionnaire is reported in Table 8. SCS scores are computed by summing the values of all 10 items in the SCS and dividing the result by 50 to normalize it in a range between 0 and 1 [34]. It is worth mentioning that this questionnaire was not part of the study reported in [21], whose experimental design we followed. However, we have chosen to introduce this questionnaire because it helps us to assess the explanations in depth. It can be objectively stated that the introduction of the questionnaire did not alter the study design, thus ensuring that our warning results are comparable to the baselines.

Table 8. SCS questionnaire items [34]

| # | Questionnaire item |
|---|---|
| 1 | I found that the data included all relevant known causal factors with sufficient precision and granularity |
| 2 | I understood the explanations within the context of my work |
| 3 | I could change the level of detail on demand |
| 4 | I did not need support to understand the explanations |
| 5 | I found the explanations helped me to understand causality |
| 6 | I was able to use the explanations with my knowledge base |
| 7 | I did not find inconsistencies between explanations |
| 8 | I think that most people would learn to understand the explanations very quickly |
| 9 | I did not need more references in the explanations: e.g., medical guidelines, regulations |
| 10 | I received the explanations in a timely and efficient manner |

### 5.3 Procedure

Participants began the study by accepting to take part through Prolific after reading the description. They were then directed to our survey by clicking on the Lime Survey link. The first page of the survey provided an overview of the study, followed by the ethical form provided by the Research Ethics Committee of King's College London after its approval by the ethical board (Ethical Clearance Reference Number: *MRA-23/24-40811*). Participants could withdraw from the study at any time; no one withdrew from the study. After accepting the information presented in the ethical form, participants were asked to follow a scenario in which they received some emails supposedly from well-known service providers, that these emails had been flagged as dangerous, and that whenever they tried to click on a link in them, a warning dialog would appear.

Each participant was then assigned to one of four experimental conditions according to a balanced Latin square design. The survey showed a web page for each of the 4 emails. Each page included a screenshot of the email at the top, followed by the sentence "Imagine that, by clicking on the link in the email, the following warning appeared:" and the corresponding warning; the same page also contained the two questionnaires (warning questionnaire and SCS). Once a participant completed all the conditions and filled a total of 8 questionnaires (2 questionnaires for each of the 4 email/warning pairs), we thanked them and directed them back to the *Prolific* platform to receive their reward. Participants were paid at a rate of £10.50/hour (above Prolific's recommended rate of £9.00/hour). We did not ask any questions that would reveal the participant's identity. The study lasted approximately 20 minutes for each participant. Prior to the study, a pilot study was conducted with 5 participants to evaluate the entire procedure; the results were not included in the analysis.

### 5.4 Data analysis

The Mann-Whitney U-Test was used to analyze the quantitative data (items 3, 4, 5, 7, and 10 of the warning evaluation questionnaire) by performing pairwise comparisons of the 8 experimental conditions, which included the 4 warnings proposed in this paper and the 4 baselines. A statistical significance level $\alpha = 0.05$ was applied. If the p-value was less than $\alpha$, the effect size was calculated [38]. The effect size *r* was categorized into three levels: small (values between 0 and 0.3), medium (values between 0.31 and 0.5), and large (values between 0.51 and 1).

We analyzed the data from the SCS questionnaire by comparing all 10 items across our four experimental conditions (W1, W2, W3, W4) using the Mann-Whitney U-Test. The same test was also applied to the overall SCS values of the 4 conditions. All the details of the inferential tests are reported in the repository under the folder "results/Warning evaluation".

We conducted an inductive thematic analysis [8] of the qualitative data from items 2, 6, and 9 to identify the major themes that would spontaneously emerge from the data. Specifically, two authors of this paper analyzed the answers to the three open questions following the 6-step thematic analysis procedure, which are data familiarization, coding, theme generation, theme review, theme naming, and theme description [8].

## 6 RESULTS

### 6.1 Quantitative results

The results of the study, which involved 320 pairwise comparisons (i.e. 8x8 conditions x 5 quantitative questionnaire items), are summarized in Figure 4. The 4 warnings proposed in this study are indicated as W1, W2, W3, and W4, while the baselines are indicated as M (manual explanation), C (Chrome), F (Firefox), and E (Edge). In this overview, we have excluded the results of the comparisons between the warnings proposed in this study as no differences were found. We have obviously also omitted the results of the baseline comparisons as they were already reported in the paper [21]. Full details of the statistical tests can be found at this link. For the remainder of this section, all the statistical differences that are reported are always in favor of the 4 warnings proposed in this study.

|    | 3. Understandability |   |   |   | 4. Familiarity |   |   |   | 5. Interest |   |   |   | 7. Risk Felt |   |   |   | 10. Trust |   |   |   |
|----|---|---|---|---|---|---|---|---|---|---|---|---|---|---|---|---|---|---|---|---|
|    | M | C | F | E | M | C | F | E | M | C | F | E | M | C | F | E | M | C | F | E |
| W1 | H | L | L | L |   |   |   |   | L | L | L | L | L |   |   |   |   | L |   |   |
| W2 | M | L | L | L |   |   |   |   | L |   |   |   |   |   |   |   | M |   |   |   |
| W3 | M | L | L | L |   |   |   |   | L |   |   |   |   |   |   |   |   |   |   |   |
| W4 | H | L | L | L |   |   |   |   | L |   | L |   |   |   |   |   | M | L |   |   |

Figure 4. Overview of statistical test results: a green cell indicates a low effect (*L*), yellow denotes a medium effect (*M*), and red indicates a high effect (*H*). **M** = manual explanation, **C** = Chrome explanation, **F** = Firefox explanation, **E** = Edge explanation. **W1** = *IP address* explanation, **W2** = *Top-Level domain mispositioned* explanation, **W3** = *URL mismatch* explanation, **W4** = *New domain* explanation.

Regarding *understandability* (item 3), all four warnings resulted in very high scores (W1 $\bar{x}$ = 4.9, W2 $\bar{x}$ = 4.75, W3 $\bar{x}$ = 4.75, W4 $\bar{x}$ = 4.9). Statistical differences emerged when compared to the four baselines: a low effect was found when comparing our warnings with Chrome (C), Firefox (F) and Edge (E), a medium effect when comparing W1/W2 with the baseline (M), while a high effect was found when comparing W1/W4 with the baseline (M).

Regarding *familiarity* (item 4), high scores were achieved for all four warnings (W1 $\bar{x}$ = 3.9, W2 $\bar{x}$ = 3.5, W3 $\bar{x}$ = 3.85, W4 $\bar{x}$ = 3.95). In contrast to the previous metric, no statistical differences emerged in any of the comparisons, including those with the baselines. Similar results were obtained for *perceived risk* (item 7), with all four warnings obtaining very high scores (W1 $\bar{x}$ = 4.2, W2 $\bar{x}$ = 4.1, W3 $\bar{x}$ = 4.0, W4 $\bar{x}$ = 4.2). However, no statistical differences were found, except for the comparison between W1 and M.

Concerning *interest* (item 5), the results were very positive, with all four warnings obtaining very low scores (the question has the opposite polarity, W1 $\bar{x}$ = 1.35, W2 $\bar{x}$ = 1.85, W3 $\bar{x}$ = 1.75, W4 $\bar{x}$ = 1.6). W1 resulted to be more interesting than all the baselines, W4 was better than F, and all our warnings were better than M. The differences had a low effect in all cases.

Finally, perceived *trust* (item 10) resulted in positive scores (W1 $\bar{x}$ = 4.25, W2 $\bar{x}$ = 4.05, W3 $\bar{x}$ = 3.8, W4 $\bar{x}$ = 4.25). Statistical differences emerged between W1 and C (low effect), W2 and M, and W4 with both C and M.

The effectiveness of the warning dialog is measured by item 8. No statistical test could be applied to measure significant differences. Nonetheless, results for this item evidently show that W1, W2, and W4 led to a higher percentage of users choosing the safest action ("to not continue to the website"), i.e., 76.2% for all the warnings. The only exception was W3, which reported a lower percentage of users (61.9%) choosing the safest action. On the contrary, all baselines led to a lower percentage of users choosing the safest action, and a higher percentage of users choosing "to be careful while continuing to the website", especially in the case of the manual explanation M (not continue = 52.8%, be careful while continuing = 42.7%).

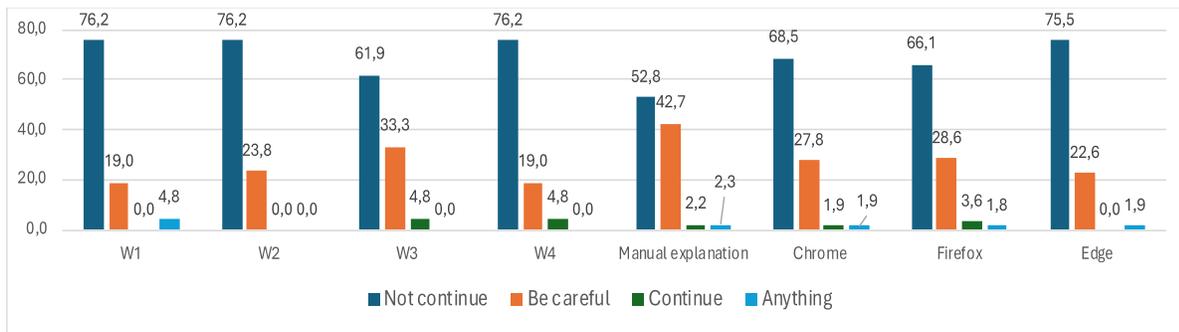

Figure 5. Percentages of each experimental condition for item 8

The last results come from the analysis of the SCS questionnaire (Figure 6), which indicates the perceived quality of the explanations through the SCS score. The score of all four warnings proposed in this study was high (W1 x̄=.835, sd=.094; W2 x̄=.804, sd=.131; W3 x̄=.791, sd=0.112; W4 x̄ =.829, sd=.098). No significant differences emerged among the different conditions for both the individual items of the SCS and for the overall SCS scores. However, scores for Item 1 ("I found that the data included all relevant known causal factors with sufficient precision and granularity") and Item 3 ("I could change the level of detail on demand") of the SCS questionnaire were lower in average (item 1 x̄=.3.925, sd=0.823; item 3 x̄=.3.4, sd=1.039).

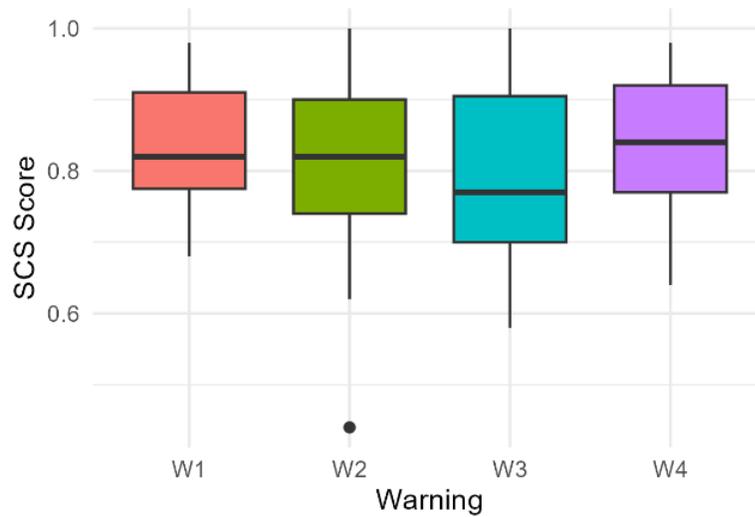

Figure 6. Box plots showing the results of the System Causality Scale questionnaire for the experimental conditions W1, W2, W3, and W4.

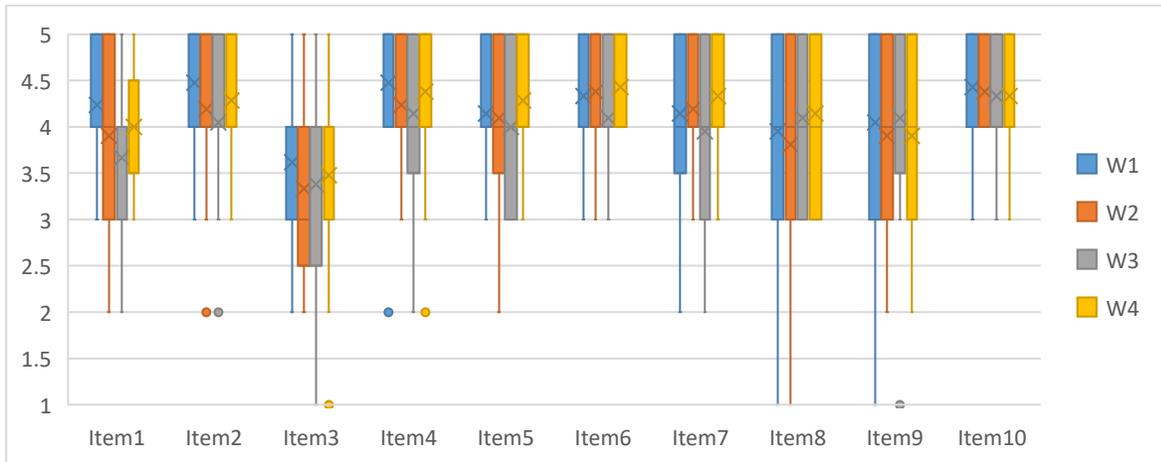

Figure 7. Box plots showing the detailed results of the 10 items of the System Causality Scale questionnaire for the experimental conditions W1, W2, W3, and W4.

**6.2 Qualitative results**

In relation to item 2 ("When you saw the warning dialog, what was your first reaction?") we identified 5 themes. The first theme, named "Emotive reaction" refers to users who reported an emotional response, such as feeling negative emotions (e.g., alert 11 times, confusion 4 times, panic 1 time, concern 8 times). The second theme, named "Stop the interaction – unsafe content", refers to users taking drastic action by interrupting the interaction due to unsafe content (26 occurrences). The third theme, named "Suspicious on the content", refers to users feeling that something dubious is happening (9 occurrences). The fourth theme, "Need to investigate", is linked to a state of uncertainty about the validity of the content, which leads users to want to investigate further before moving on (5 occurrences). The last theme, called "Nothing", relates to the lack of user reaction even after being exposed to a warning.

The analysis of the answers to item 9 ("What do you think this warning dialog means?") led to identifying three themes. The most common theme, named *"Generic Danger"*, suggests that users interpreted the warning as referring to a potential threat to their personal information (26 occurrences), as well as to their devices (2 times), or a generic danger (10 times). The second theme, *"Phishing content"*, suggests that users perceive a definite risk of phishing. Within this theme, users perceived as dangerous the website to be opened (26 times), the email (2 times) or the link itself (5 times). The third and final theme called *"Potential unsafe content"* is similar to the previous one; however, it differs in the degree of certainty of the users. In this theme, in fact, we find responses from users who are unsure whether the content is phishing. The content perceived as potentially unsafe is, again, differentiated between the email (3 times), link (2 times) and website (11 times). Figure 8 provides a summary of the themes and their distribution within the four warnings.

Finally, the results for item 6 report that some words in the explanations were reported as either confusing or too technical for the user. These words include "domain" (reported 7 times), "top-level" (5 times), "deceptive" (2 times), "IP address" (1 time), "very new" (1 time – referring to the domain), "string of numbers" (1 time).

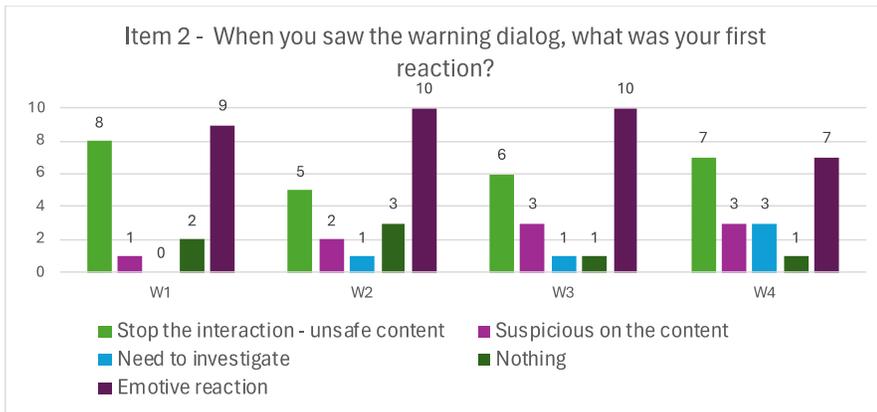

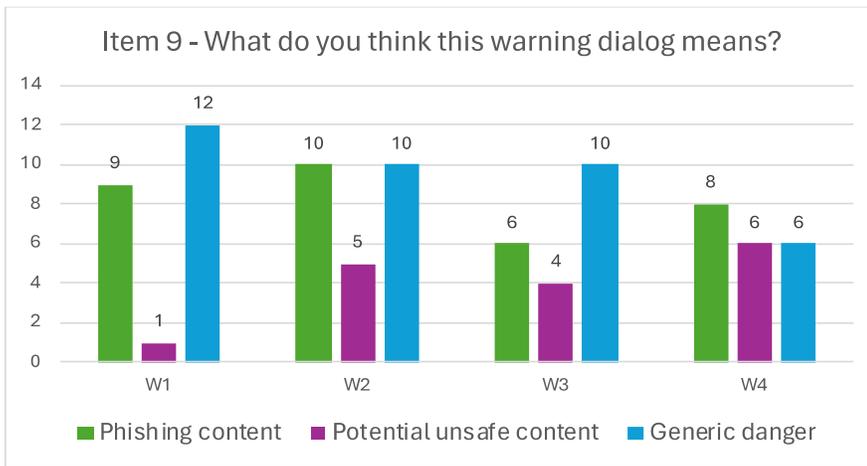

Figure 8. Distribution of the codes of the themes for each warning dialog for item 2 (a) and item 9 (b)

### 6.3 Limitations

This study has some limitations which are described in the following. First of all, the explorative nature of the study cannot constitute a generalization of results, but rather suggests that applications of LLMs in high-stakes contexts like phishing defense is feasible, at least to warn the users about the risks of malicious emails. Moreover, we have to consider that the explanation messages that are produced by an LLM are highly subject to the prompt used to generate them. At the same time, we used a baseline for our comparison that included warnings that were manually produced by experts [21]; the quality of these warnings might depend on the designers' skills.

The explanation messages that were evaluated in this study were produced with APOLLO for 4 specific emails, each with one specific phishing feature. The choice of the emails to show to the participants and of the feature to explain has a great impact on the external validity of the results. Moreover, the features that we used in this study were different from those used for the explanations in the baseline [21].

The SCS questionnaire has not yet been applied enough in the literature to compute a percentile ranking, unlike its counterpart for measuring usability, the System Usability Scale, on which it was inspired [11]. Therefore, the SCS scores of this study can only be considered as absolute values, hopefully constituting a baseline for comparison in future studies.

## 7 DISCUSSIONS AND LESSONS LEARNED

The aim of this exploratory study was to answer the RQ: "*How do users perceive warning dialogs for phishing attacks generated by LLMs?*". We attempted to distill some answers to the research question in the form of lessons learned, which are reported in the following.

**LLM-generated explanations show promise in protecting users.** The analysis of item 8 reveals that no users would engage in the most dangerous and unsafe behavior of continuing to the website. In fact, almost all participants chose the safest actions, i.e., "to not continue to the website" (76.2% of the time) and "be careful while continuing to the website" (19% of the time). These actions resulted in higher percentages than the baselines; notably, in the manual warning the safest action ("Not continue") was chosen by only 58% of participants. Moreover, this very promising result is supported by the high levels of trust (item 10) scored by the LLM-generated warnings, which was higher than that of Chrome for W1 and W4, and much higher than that of the manual explanation in the case of W2 and W4. The lack of trust by users is indeed a weakness of current warnings [21] and improving it could enhance the warnings' effectiveness.

**LLM-generated explanations are perceived as very understandable, interesting and of high quality**. This study revealed that explanations generated with APOLLO were, with a low effect, more understandable (item 3) and interesting (item 5) than those of Chrome, Firefox, and Edge, despite the presence of the additional explanation message that could potentially overload the users. LLM-generated warnings also outperformed the manually written explanation with a medium (W2, W3) or high (W1, W4) effect. This finding is very significant, considering that the manual baseline was generated following a rigorous and time-consuming process driven by metrics that assess text comprehensibility, readability, and sentiment [21]. Additionally, using LLMs to generate explanations has been shown to increase user interest: in this study, all four warnings scored very positively for interest, with W1 outperforming all the baselines, W4 outperforming Firefox, and all the warnings outperforming the manual explanation. The aspect of interest is crucial because, if it is low, the importance that users place on the warning is reduced, leading to the warning recommendations being substantially ignored. Moreover, the positive results of the SCS questionnaire indicate that LLM-generated explanations are not only understandable and interesting, but also perceived as high-quality overall. In fact, all of our warnings received an SCS score of 0.79 (for W3) out of 1 or higher.

**Integrating the need for cognition into warnings.** The evaluation of warnings carried out through the analysis of the System Causality Scale questionnaire items has yielded valuable insights. In particular, responses to Item 1 ("I found that the data included all relevant known causal factors with sufficient precision and granularity") and Item 3 ("I could change the level of detail on demand") indicated that users may perceive a lack of granularity in the warnings and may prefer more detailed explanations on demand. These findings are closely related to the concept of Need for Cognition (NFC)[14], which has significant implications for the design of effective warnings. Users with high NFC are not particularly satisfied with superficial explanations and seek detailed causal factors to comprehend why an email might be a scam fully, and vice versa. Warnings should, therefore, provide more precise and granular information to cater to users' cognitive needs, enhancing their understanding and trust in the system. Implementing a customizable level of detail in warnings can accommodate both high and low NFC users.

**LLM-generated explanations offer good support for identifying false positives**. When designing warnings for phishing attacks, it is crucial to help users make informed decisions not only in the case of a genuine phishing email but also in the case of false positives. Neglecting this aspect can compromise the user's productivity, as they could miss important emails for an erroneous classification. Unfortunately, this aspect is often overlooked in the evaluation of warning dialogs (e.g., [54]). This study considered this aspect by exposing the participants to a legitimate email for which we made APOLLO generate an erroneous, yet credible, explanation (included in W3) to simulate the case of a false positive. The results showed that W3 received the lowest trust score, indicating that users were able to understand that the warning may be incorrect due to the explanation provided. This interesting finding is supported by the analysis of item 9, which assessed the user's understanding of the warning dialog: it is worth noting that W3 received the lowest score for "Phishing content" (Figure 8b), meaning that users perceived the email linked to W3 as less dangerous than the other ones. Moreover, although no significant differences emerged in the SCS scores, it is interesting to observe that W3 resulted to be of lower quality compared to the other warnings, on average (Figure 6). This might indicate that explanations attempting to explain a false positive could correctly constitute a red flag for the user.

**The balance between rational and emotional thinking needs to be investigated**. During the analysis of item 2, which pertains to users' reactions, it was found that some users applied emotional thinking (theme "Emotive reaction"), while others used rational thinking (the remaining themes). Distinguishing between emotional and rational thinking in this context is crucial as it may affect the response that users have. Cyberattacks can often trigger emotional responses such as fear or panic, as some participants have noted. It is, therefore, important to understand how users react emotionally to warning dialogs in order to design messages that alleviate fear and encourage a constructive response. On the other hand, users also evaluate the information presented in the warning dialog in a rational manner; this includes assessing potential risks, understanding consequences, and making informed decisions about how to proceed, also depending on their expertise [9]. Designing effective warnings requires finding the right balance between the two extremes, also considering user profiles.

## 8 CONCLUSIONS AND FUTURE WORK

In this paper, we presented APOLLO, a tool that uses GPT-4o to classify emails and explain why an email could be a phishing email. A thorough evaluation was performed to assess the tool's performance in classifying emails as phishing/genuine, covering several cases. The results demonstrate very high accuracy in detecting malicious emails even in challenging scenarios (i.e., when the LLM is primed with wrong external information about the URL). Furthermore, we validated APOLLO by conducting a user study to evaluate the generated warning messages with explanations in the context of phishing attacks. The results were compared to different baselines [21], and LLM-generated warnings resulted in being significantly more understandable, interesting, and trustworthy than manually generated ones. Moreover, the SCS scores, which indicate the overall quality of explanations and their suitability for the intended purpose [34], were high for all the LLM-generated warnings. These results are very promising and suggest that we should favor an LLM-powered approach when generating warnings for phishing emails, as it would also improve the scalability, and efficiency of the design process.

Explanations generated by an LLM using prompt engineering are different from those generated by other more traditional methods (like eXplainable AI) and surely have several drawbacks [7, 12, 44, 61]; e.g., LLMs may suffer from hallucinations, the generated explanations may not reflect the internal processes of the model, etc. Nevertheless, in this study, we have highlighted several qualities that LLM-generated text tends to have (when the model is correctly prompted); moreover, APOLLO represents a proof-of-concept of a unified system for both classifying phishing emails and generating warnings to alert the user. Such an approach maximizes the chances of protection against phishing attacks, as it provides both technological and human defenses. In this work, we employed prompt engineering and iterative manual refinement to design prompts that could fulfill our classification and explanations generation tasks, as it is currently the most widespread technique for developing LLM applications. In future work, automatic or semi-automatic solutions like PE2 [73], Azure prompt flow[9], and flow engineering [56] will be considered to create alternative prompts, allowing us to assess the APOLLO classification performances with alternative prompts.

The evaluation study reported in this paper solely involved GPT-4o by OpenAI. This choice was driven by considering that this model is one of the most performant LLM in classification tasks (at the time of writing this article), generally carrying superior performance in text evaluation tasks compared to Claude3 Opus, Gemini Pro 1.5, Gemini Ultra 1.0 and Llama3 400b [52]. However, it would be useful to perform a further evaluation study to investigate how different LLMs behave in the phishing emails classification tasks, especially when exposed to external URL information. Moreover, given the proliferation of LLMs, a Mixture-of-Agents [66] approach could harness the collective expertise of multiple individual models and might lead to superior performance. Therefore, it would be interesting to include these evaluations as part of future studies.

The user study reported in this article will lay the foundation for a controlled experiment that will measure the actual effectiveness of the LLM-generated warnings in protecting users by means of metrics such as the click-through rate of users when exposed to the warning in a simulated scenario, in a setting similar to studies like [13, 54]. This future study will also serve as a means to compare LLM-generated warnings to manually generated messages and assess whether there is any difference between the two in users' protection. Compared to the present survey, which involved a limited number of participants, the second study will be extended to include many more participants to have more statistical power in the results. This study can also include the assessment of LLM classification variability on the consistency and reliability of user interactions. This may require conducting multiple trials with varying inputs and carefully controlling for confounding variables to isolate the effects of LLM integration on user behavior.

From the results of the thematic analysis, we have highlighted a difference between emotional and rational thinking, which will be further investigated as influential factors in phishing susceptibility. Tools such as biometrics sensors will be employed in a future study for measuring emotional user data in a laboratory setting. Finally, since the technological advancement of LLMs has made it possible to process images, we will use state-of-the-art multi-modal models such as GPT-4o to also process an email visually; this approach might effectively increase the model's detection capabilities. Furthermore, multi-modal LLMs may be used to generate not only warning messages, but the entire warning dialog, thus increasing the polymorphic capabilities of the system; e.g., a warning generated in this way could include visual phishing cues in the email that might be more relevant to the user. Warnings could also be tailored according to different characteristics of the user, such as their expertise level; in fact, users with different levels of knowledge might benefit from different types of explanations [49, 67], resulting in more effective protection overall.

**Acknowledgments**

This work has been supported by the Italian Ministry of University and Research (MUR) and by the European Union-NextGenerationEU, under grant PRIN 2022 PNRR "DAMOCLES: Detection And Mitigation Of Cyber attacks that exploit human vuLnerabilitiES" (Grant P2022FXP5B) CUP: H53D23008140001. This work is partially supported by the co-funding of the European Union - Next Generation EU: NRRP Initiative, Mission 4, Component 2, Investment 1.3 - Partnerships extended to universities, research centres, companies and research D.D. MUR n. 341 del 5.03.2022 - Next Generation EU (PE0000014 – "Security and Rights In the CyberSpace – SERICS" - CUP: H93C22000620001). The research of Francesco Greco is funded by a PhD fellowship within the framework of the Italian "D.M. n. 352, April 9, 2022"- under the National Recovery and Resilience Plan, Mission 4, Component 2, Investment 3.3 - PhD Project

---

[9] https://learn.microsoft.com/en-us/azure/machine-learning/prompt-flow/overview-what-is-prompt-flow

"Investigating XAI techniques to help user defend from phishing attacks", co-supported by "Auriga S.p.A." (CUP H91I22000410007).

# APPENDIX

## A. Post-hoc tests to highlight differences between predicted labels (chi-square) and predicted probabilities (Tukey's HSD)

Table 9. Post-hoc tests performed with chi-square and Tukey's Honestly Significant Difference (HSD) test.

|  | Chi-square | | Tukey's HSD |
| --- | --- | --- | --- |
| **Conditions** | **Statistic** | **p-value** | **p-value** |
| noURL vs Q0 | 32.531 | **.000** | **.000** |
| noURL vs Q25 | 0.0374 | **.000** | .353 |
| noURL vs Q50 | 0.0374 | **.000** | .283 |
| noURL vs Q75 | 0.0374 | **.000** | .811 |
| noURL vs Q100 | 0.0374 | **.000** | .876 |
| Q0 vs Q25 | 53.409 | **.000** | **.000** |
| Q0 vs Q50 | 53.409 | **.000** | **.000** |
| Q0 vs Q75 | 53.409 | **.000** | **.000** |
| Q0 vs Q100 | 53.409 | **.000** | **.000** |
| Q25 vs Q50 | 1 | 1 | 1 |
| Q25 vs Q75 | 1 | 1 | .978 |
| Q25 vs Q100 | 1 | 1 | .954 |
| Q50 vs Q75 | 1 | 1 | .956 |
| Q50 vs Q100 | 1 | 1 | .920 |
| Q75 vs Q100 | 1 | 1 | 1 |
| Q0 vs Q25ERR | 2378.814 | **.000** | **.000** |
| Q0 vs Q50ERR | 2415.25 | **.000** | **.000** |
| Q0 vs Q75ERR | 2387.898 | **.000** | **.000** |
| Q0 vs Q100ERR | 2373.372 | **.000** | **.000** |
| Q25ERR vs Q50ERR | 0.182 | 1 | .732 |
| Q25ERR vs Q75ERR | 0.000 | 1 | .845 |
| Q25ERR vs Q100ERR | 0.000 | 1 | **.000** |
| Q50ERR vs Q75ERR | 0.090 | 1 | .243 |
| Q50ERR vs Q100ERR | 0.244 | 1 | **.000** |
| Q75ERR vs Q100ERR | 0.024 | 1 | **.000** |

**B. Screenshots of the phishing emails and the false positive email used in the study**

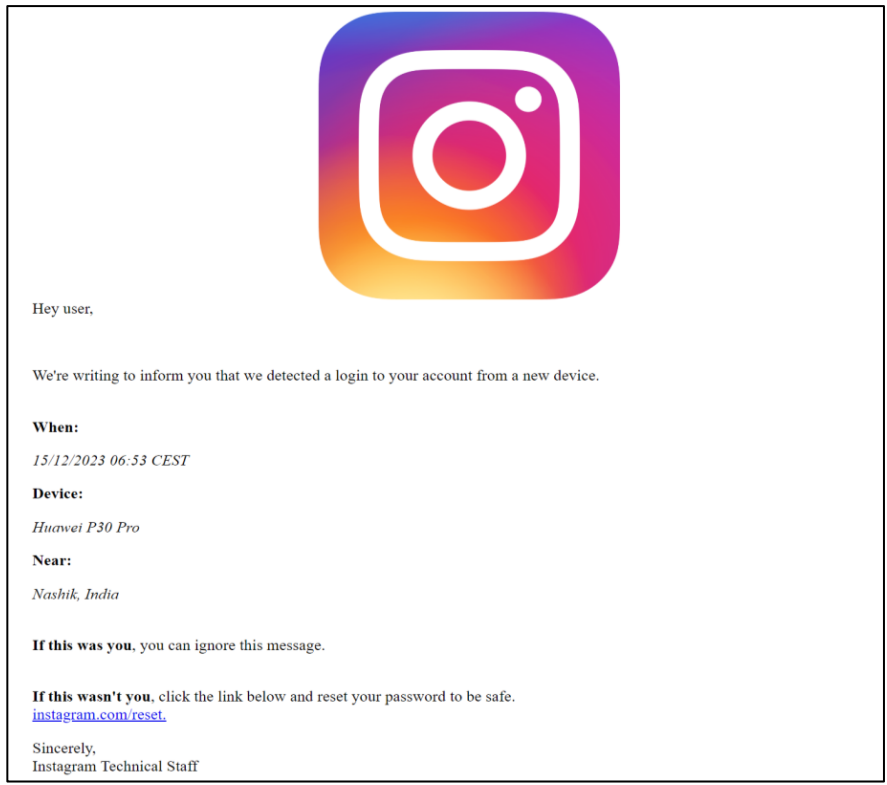

Figure 9. Email associated with warning W1 (IP Address feature)

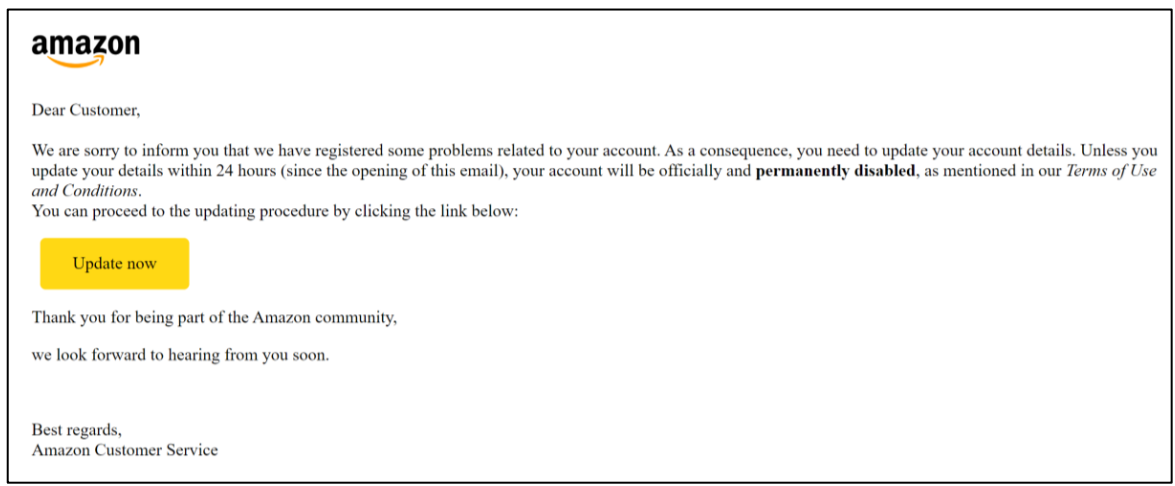

Figure 10. Email associated with warning W2 (Top-level domain mispositioned)

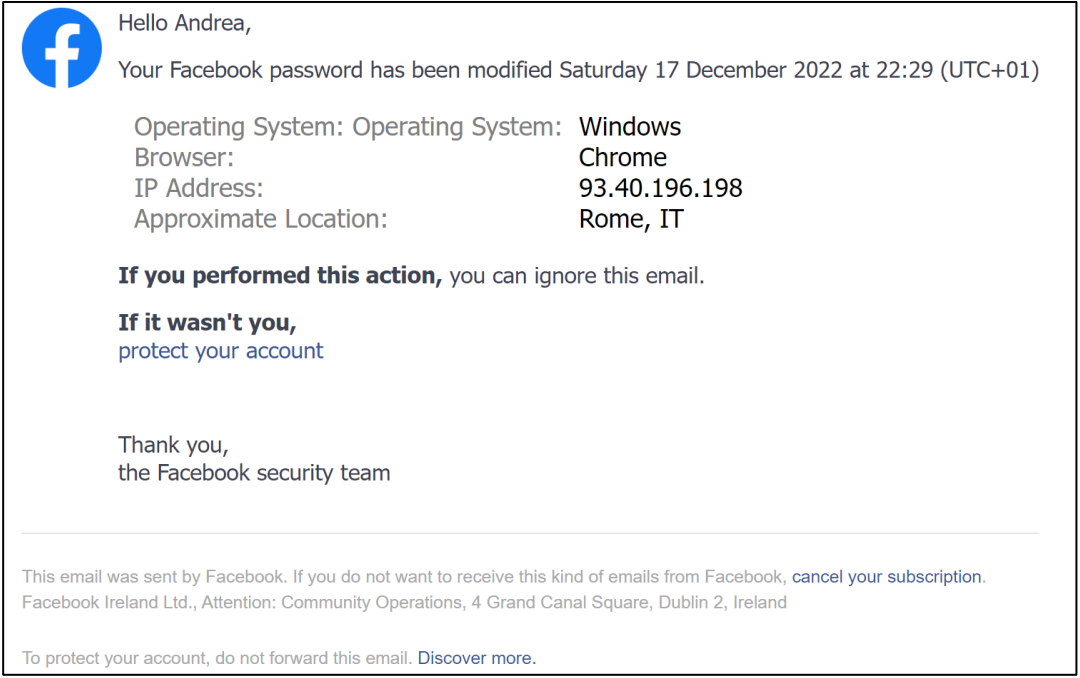

Figure 11. Email associated with warning W3 - false positive (link mismatch)

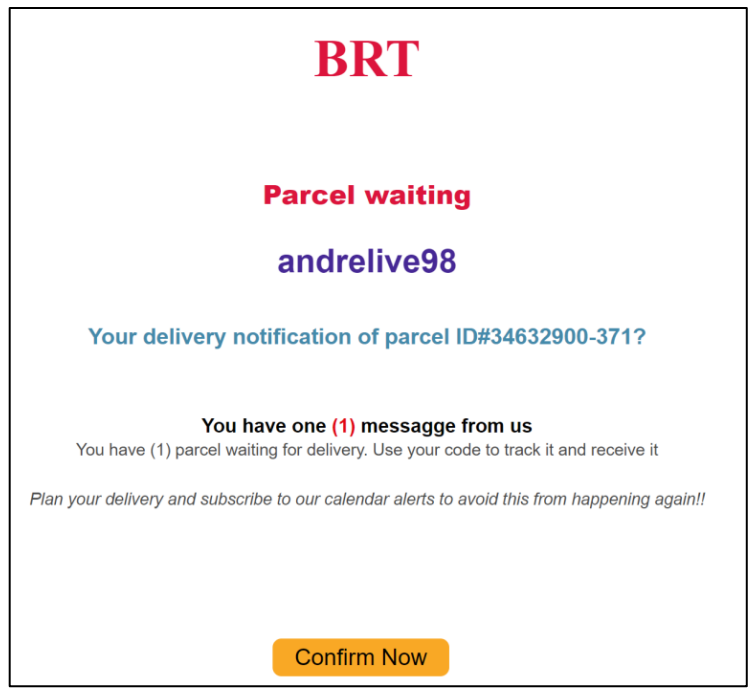

Figure 12. Email associated with warning W4 (newly created domain)